\documentclass[11pt]{article}

\usepackage[english]{babel}

\usepackage[a4paper,top=2cm,bottom=2cm,left=2.5cm,right=2.5cm,marginparwidth=1.75cm]{geometry}
\usepackage{setspace}
\usepackage[colorlinks=true, allcolors=blue]{hyperref}
\usepackage[utf8]{inputenc}
\usepackage[T1]{fontenc}
\usepackage{amsmath,amssymb}
\usepackage[numbers]{natbib}
\usepackage{hyperref}
\usepackage{titlesec}
\usepackage{caption}
\usepackage{setspace}
\usepackage{multicol}
\usepackage{ragged2e} 
\usepackage[dvipsnames]{xcolor} 
\usepackage{tcolorbox} 

\title{Why the Northern Hemisphere Needs a 30–40\,m Telescope and the Science at Stake:\\ 
Ultra--Low-Mass Dwarf Galaxies Across the Boreal Cosmic Web}
\author{
J. Alfonso L. Aguerri$^{*,1,2}$, 
Jes\'us Falc\'on-Barroso$^{1,2}$, 
Mar\'ia Argudo-Fern\'andez$^{3}$,\\ 
Yago Ascasibar$^{4}$,
Marc Balcells$^{1,2}$,
Bahar Bidaran$^{3}$,
Virginia Cuomo$^{5}$,\\ 
David Fern\'andez-Arenas$^{6,7}$,
Anna Ferr\'e-Mateu$^{1,2}$,
Rub\'en Garc\'ia-Benito$^{8}$,\\
Rosa María Gonz\'alez Delgado$^{8}$,
Marcella Longhetti$^{9}$,
Pavel Mancera-Pi\~na$^{10}$\\
Antonino Marasco$^{11}$,
Lorenzo Morelli$^{12}$,
Reynier F. Peletier$^{13}$,\\
Isabel P\'erez Mart\'in$^{3}$, 
Francesca Pinna$^{1,2}$,
Daniel Rosa Gonz\'alez$^{14}$,\\
Marc Sarzi$^{15}$,
Alexandre Vazdekis$^{1,2}$, 
Marc Verheijen$^{13}$,\\
Pedro Villalba Gonz\'alez$^{16}$,
Anne-Marie Weijmans$^{17}$,
Stefano Zarattini$^{18}$\\[6pt]
{\small \texttt{$^{*}$jalfonso@iac.es}}\\\\
{\small $^{1}$Instituto de Astrof\'isica de Canarias (Spain)}\\
{\small $^{2}$Departamento de Astrof\'isica. Universidad de La Laguna (Spain)}\\
{\small $^{3}$Departamento de F\'isica Te\'orica y del Cosmos, Universidad de Granada (Spain)}\\
{\small $^{4}$Departamento de F\'isica Te\'orica, Universidad Aut\'onoma de Madrid (Spain)}\\
{\small $^{5}$Departamento de Astronomía, Universidad de La Serena, (Chile)}\\
{\small $^{6}$Canada-France-Hawaii Telescope (United States)}\\
{\small $^{7}$Planetarium La Enseñanza, Medellín, Antioquia. (Argentina)}\\
{\small $^{8}$Instituto de Astrofisica de Andalucía (Spain)}\\
{\small $^{9}$INAF – Astronomical Observatory of Brera (Italy)}\\
{\small $^{10}$Leiden Observatory (the Netherlands)}\\
{\small $^{11}$INAF – Padova Astronomical Observatory (Italy)}\\
{\small $^{12}$Instituto de Astronom\'{\i}a y Ciencias Planetarias, Universidad de Atacama (Chile)}\\
{\small $^{13}$Kapteyn Astronomical Institute, University of Groningen (the Netherlands)}\\
{\small $^{14}$Instituto Nacional de Astrof\'isica, \'Optica y Electr\'onica (Mexico)}\\
{\small $^{15}$Armagh Observatory (United Kingdom)}\\
{\small $^{16}$University of British Columbia (Canada)}\\
{\small $^{17}$University of StAndrews (United Kingdom)}\\
{\small $^{18}$Centro de Estudios de F\'{\i}sica del Cosmos de Arag\'on (Spain)}
}

\begin{document}
\maketitle
\newpage

\section{Scientific context}

Dense clusters serve as powerful laboratories for studying environmental processing. The Coma Cluster -- one of the densest nearby systems with no counterpart in the Southern Hemisphere -- is exceptionally rich in dwarf galaxies and exhibits an extreme environment in which tidal interactions, galaxy harassment, and ram-pressure stripping operate with high efficiency \citep[][]{Dressler1980,Moore1998,BoselliGavazzi2006}. With its high velocity dispersion, deep gravitational potential, and dense intracluster medium, Coma represents a regime where multiple environmental mechanisms act simultaneously, often on short timescales. Integral-field spectroscopic studies of Coma dwarfs have revealed detailed internal kinematics, spatially resolved stellar-population gradients, and clear signatures of transformation-driven substructures, including embedded discs, bars, spiral features, and kinematically decoupled components \citep{deRijcke2003,deRijcke2004,Chilingarian2007,Thomas2006}. These signatures indicate that many dwarfs retain memory of their progenitor morphologies or accretion histories, even after significant environmental processing. Such morphological and dynamical features provide direct clues to their origins---whether they are long-lived cluster members that formed early, recently accreted late-type galaxies currently undergoing transformation, or products of tidal interactions such as harassment or tidal stirring \citep{Conselice2001,Conselice2003}. Understanding which pathways dominate requires large, homogeneous spectroscopic samples with sufficient spatial resolution to map internal structure and detect low-surface-brightness components.\looseness-2\medskip

While Coma exemplifies the most extreme end of environmental influence, dwarf ellipticals and other low-luminosity systems dominate the number counts in rich clusters \citep{Sandage1985,BinggeliJerjen1998}, and their structural and kinematic properties show a strong dependence on local environment \citep{Fillingham2018,RomeroGomez2023}. The prevalence of quenched, pressure-stripped, and dynamically heated dwarfs toward cluster centers highlights the cumulative impact of dense environments on low-mass galaxy evolution. However, developing a complete picture of dwarf-galaxy evolution requires sampling across the full hierarchy of environments. Filaments, galaxy groups, and the outskirts of massive structures---including those in the Northern Hemisphere such as Virgo, Coma, and Pisces--Perseus---provide intermediate-density regimes where preprocessing, backsplash orbits, and gentle stripping can be disentangled from the more violent mechanisms acting in cluster cores \citep{Wetzel2015,Laigle2015}. These regions host galaxies at different evolutionary stages, from infalling systems just beginning to experience environmental influence, to backsplash dwarfs that have already passed through cluster environments and now reside at large radii. Mapping how internal kinematics, gas content, and stellar populations vary across these environments is essential for reconstructing the relative importance and timescales of environmental processes.\looseness-2\medskip

By observing matched stellar-mass dwarf galaxies in Coma, Virgo, infalling groups, filaments, and low-density regions, a northern 30--40\,m telescope would enable a systematic, controlled comparison of how star formation, chemical enrichment, and mass assembly depend on both local density and position within the cosmic web \citep{Haynes1988,Kubo2007}. Such a facility would, for the first time, allow spatially resolved spectroscopy of ultra--low-surface-brightness dwarfs over a representative range of environments, providing dynamical mass estimates, orbital histories, and chemo-dynamical diagnostics for thousands of systems. Isolated dwarfs in underdense regions serve as essential baselines---representing minimally perturbed evolutionary pathways against which environmentally processed systems can be directly evaluated \citep{Weisz2014}. Combining these baselines with observations of environmentally transformed dwarfs will yield stringent constraints on environmental quenching efficiency, the role of dark-matter halo structure in shaping dwarf evolution, and the balance between internal and external drivers of galaxy transformation.\looseness-2

\section{Science Challenge}

Ultra-low-mass galaxies with stellar masses between 10$^{5}$ and 10$^{7}$\,M$_\odot$ dominate galaxy number counts and are central to understanding galaxy formation and evolution. Their shallow potentials amplify the effects of stellar feedback, reionization, and stochastic chemical enrichment \citep{Weisz2014}, while their internal dynamics directly probe the nature of dark matter \citep{Bullock2017}. The lowest-mass dwarfs retain signatures of early star formation and reionization \citep{Jeon2017}, and are sensitive to alternatives to cold dark matter such as warm or self-interacting dark matter \citep{Lovell2014,Vogelsberger2016}. They therefore connect astrophysics to fundamental cosmology.\medskip

Yet despite their importance, the faintest dwarfs remain largely unexplored spectroscopically. Even in the Local Volume, only a subset of systems have velocity dispersions, metallicity distributions, or spatially resolved stellar-population gradients measured with sufficient precision to confront theoretical models. At distances beyond $\sim$5--10\,Mpc, current 8--10\,m facilities reach only the brightest dwarfs in Virgo and Fornax, and cannot resolve internal kinematics or stellar-population gradients in Coma-like environments \citep{Toloba2014}. This gap severely limits our ability to assess the relative roles of internal versus external processes, and to quantify the diversity of dark-matter halos at the lowest masses. However, nowadays intruments and facilitie shave transformed the field: they have revealed the diversity of dwarf-galaxy populations in clusters and groups, uncovered the role of environmental quenching, and provided the first detailed measurements of internal kinematics and chemical enrichment in systems down to $M_\star \sim 10^{7-8}\,M_\odot$. These advances have clarified key questions---such as the efficiency of ram-pressure stripping, the timescales of preprocessing, and the connection between dark-matter halos and star-formation histories---yet they also highlight a fundamental limitation: the faintest and most weakly bound dwarfs, those with $M_\star < 10^{5-7}\,M_\odot$, remain essentially inaccessible beyond the Local Group. Ultra--low--mass dwarf galaxies will be prime targets in the Local Group for deep resolved stellar--population studies with the ELT. {\it Crucially, these same low--mass systems can also be studied in integrated light out to Virgo and Coma, allowing 30--40\,m telescopes to link resolved and unresolved views of dwarf--galaxy evolution within a unified framework.}\medskip

Achieving robust measurements of dark-matter-dominated systems across different environments requires the leap in sensitivity, spatial resolution, and multiplexing provided by a 30--40\,m aperture. Such a facility offers transformative capabilities: it allows spatially resolved spectroscopy of dwarfs an order of magnitude fainter than currently possible; it enables chemo-dynamical mapping of systems with surface brightnesses $\mu_r > 25$--27\,mag\,arcsec$^{-2}$; and it provides the depth needed to probe dwarfs embedded in the dense intracluster medium of Coma or in the filamentary structures feeding the Pisces--Perseus supercluster. This ultra--low--mass regime is crucial because it is where feedback becomes extreme, where reionization is expected to leave the strongest signatures, and where environmental effects operate with maximum efficiency. Understanding how these galaxies form, survive, or are disrupted is essential for testing galaxy-formation models, constraining dark-matter physics, and completing the empirical picture of quenching across the cosmic web.\medskip

The northern hemisphere contains unique structures spanning the full environmental spectrum: Virgo, Coma, and the Pisces--Perseus supercluster \citep{Haynes1988}. Together they provide an unparalleled laboratory in which to trace dwarf-galaxy evolution from dense cluster cores to infalling groups, filaments, and isolated void-like regions. The ability to construct matched stellar-mass samples across these environments allows a direct, controlled measurement of environmental effects, overcoming current limitations imposed by sample inhomogeneity and observational bias. This science case also benefits from strong synergies with ongoing and future northern surveys (DESI, SDSS, WEAVE, CAVITY, Euclid, HST, JWST). These facilities will deliver exquisitely deep imaging, high-precision distances, statistical redshift samples, and resolved stellar populations. A 30--40\,m telescope provides the crucial spectroscopic follow-up required to transform these photometric detections into physically interpretable constraints on dark matter, star formation, and chemical enrichment.\medskip

Combining a wide environmental view of ultra-low-mass dwarfs across the northern sky yields a coherent set of ambitious science goals that extend beyond traditional galaxy-evolution studies:

\begin{itemize}
    \item \textbf{Map the environmental dependence of dwarf-galaxy properties.} Measure ages, metallicities, abundance patterns, and star-formation histories for dwarfs across cluster cores, infalling groups, filaments, and low-density regions, and quantify how these properties vary with environment at fixed stellar mass. A 30-meter–class telescope is required because only its resolving power can access few-parsec physical scales that current facilities cannot reach, even in much closer systems such as Virgo or Fornax. Such resolution also enables detection of the rare, extremely luminous tracers of underlying stellar populations (globular and super star clusters, planetary nebulae, and the brightest supergiants), which remain inaccessible for dwarf galaxies beyond the Local Group and especially for the ultra–low-mass regime.
    \item \textbf{Identify and characterise internal substructures.} Use spatially resolved kinematics and line-strength maps to detect embedded discs, bars, spiral features, and kinematically decoupled components in dwarfs, and to relate them to formation and transformation channels.
    \item \textbf{Constrain mass distributions and dark-matter content.} Derive dynamical mass profiles for dwarfs, including dark-matter fractions within effective radii, and test whether cluster dwarfs occupy the same scaling relations (e.g.\ Fundamental Plane analogues) as their more massive counterparts.
    \item \textbf{Disentangle formation pathways in dense regions.} In Coma, determine whether dwarfs are primordial cluster members, transformed late-type galaxies that have been quenched and heated, or products of tidal interactions and mergers. Compare their properties to dwarfs in less extreme environments.
    \item \textbf{Connect local dwarfs to cosmological models.} Use the observed distribution of structural, kinematic, and stellar-population properties to test predictions of hierarchical galaxy-formation models and to constrain the role of feedback, reionization, and environmental quenching at the lowest masses.
\end{itemize}

\section{Capability Requirements for Solving the Challenge}

This program aims to constrain the mass distributions and dark-matter content of dwarf galaxies by deriving dynamical mass profiles, measuring dark-matter fractions within effective radii, and testing whether cluster dwarfs follow scaling relations analogous to the Fundamental Plane; to disentangle dwarf-galaxy formation pathways in dense environments by distinguishing primordial cluster members from transformed late-type systems or products of tidal interactions and mergers and comparing them with dwarfs in lower-density environments; and to connect local dwarf populations to cosmological models by using their structural, kinematic, and stellar-population properties to test hierarchical galaxy-formation predictions and constrain the roles of feedback, reionization, and environmental quenching at the lowest masses. Ultra–low–mass dwarfs ($10^5-10^7\,M_\odot$) are extremely compact and faint, with effective radii of $R_{\rm e} \sim 50$--$300$\,pc ($0.2''$--$0.8''$  at Virgo and Coma), mean surface brightnesses $\mu_r \sim 24$--$27~\mathrm{mag~arcsec^{-2}}$, colors $(g-r)\sim 0.3$--$0.8$, and stellar velocity dispersions of only 5–20 km s$^{-1}$, placing them beyond the practical limits of current 8–10 m telescopes for resolved or high-S/N spectroscopy beyond the Local Group. Meeting these goals requires a deployable, multi-object integral-field spectrograph covering roughly 3500--9000\,\AA\ at $R\sim5000$, with IFUs of 1$\times$1 arcsec and sub-arcsecond spatial sampling to encompass dwarf effective radii at Coma distances, combined with high multiplexing over a wide field matched to dwarf surface densities from one to several per arcmin$^2$ to enable efficient surveys of tens to hundreds of galaxies across diverse environments. All this should be assisted with AO capabilities.

\begin{tcolorbox}[colback=RoyalBlue!5!white,colframe=black!75!black, width=\textwidth]
{\it The resulting dataset will provide an essential bridge between near-field cosmology and galaxy-formation theory, establishing dwarf galaxies over the entire range of masses as precision laboratories for both astrophysics and fundamental physics.}
\end{tcolorbox}

\begin{multicols}{2}
\renewcommand{\bibfont}{\fontsize{8pt}{10pt}\selectfont}
\bibliographystyle{aa}
\bibliography{refs}

@ARTICLE{Toloba2014,
       author = {{Toloba}, E. and {Guhathakurta}, P. and {Peletier}, R.~F. and {et al.}},
      journal = {\apjs},
         year = 2014,
       volume = {215},
       number = {2},
          eid = {17},
        pages = {17},
}

@article{Haynes1988,
       author = {{Haynes}, Martha P. and {Giovanelli}, Riccardo and {Starosta}, Boris M. and {Magri}, Christopher},
        title = "{A 21 CM Survey of the Pisces-Perseus Supercluster. III. The Region North of +38 Degrees}",
      journal = {\aj},
         year = 1988,
        month = mar,
       volume = {95},
        pages = {607},
          doi = {10.1086/114660},
       adsurl = {https://ui.adsabs.harvard.edu/abs/1988AJ.....95..607H}
}

@ARTICLE{Dressler1980,
       author = {{Dressler}, A.},
        title = "{Galaxy morphology in rich clusters: implications for the formation and evolution of galaxies.}",
      journal = {\apj},
         year = 1980,
        month = mar,
       volume = {236},
        pages = {351-365},
          doi = {10.1086/157753},
       adsurl = {https://ui.adsabs.harvard.edu/abs/1980ApJ...236..351D}
}

@article{Moore1998,
       author = {{Moore}, Ben and {Lake}, George and {Katz}, Neal},
        title = "{Morphological Transformation from Galaxy Harassment}",
      journal = {\apj},
         year = 1998,
        month = mar,
       volume = {495},
       number = {1},
        pages = {139-151},
          doi = {10.1086/305264},
archivePrefix = {arXiv},
       eprint = {astro-ph/9701211},
 primaryClass = {astro-ph},
       adsurl = {https://ui.adsabs.harvard.edu/abs/1998ApJ...495..139M}
}

@article{BoselliGavazzi2006,
       author = {{Boselli}, Alessandro and {Gavazzi}, Giuseppe},
        title = "{Environmental Effects on Late-Type Galaxies in Nearby Clusters}",
      journal = {\pasp},
         year = 2006,
        month = apr,
       volume = {118},
       number = {842},
        pages = {517-559},
          doi = {10.1086/500691},
archivePrefix = {arXiv},
       eprint = {astro-ph/0601108},
 primaryClass = {astro-ph},
       adsurl = {https://ui.adsabs.harvard.edu/abs/2006PASP..118..517B}
}

@article{deRijcke2003,
       author = {{De Rijcke}, S. and {Dejonghe}, H. and {Zeilinger}, W.~W. and {Hau}, G.~K.~T.},
        title = "{Embedded disks in Fornax dwarf elliptical galaxies}",
      journal = {\aap},
         year = 2003,
        month = mar,
       volume = {400},
        pages = {119-125},
          doi = {10.1051/0004-6361:20021866},
archivePrefix = {arXiv},
       eprint = {astro-ph/0301524},
 primaryClass = {astro-ph},
       adsurl = {https://ui.adsabs.harvard.edu/abs/2003A&A...400..119D}
}

@article{deRijcke2004,
       author = {{De Rijcke}, S. and {Dejonghe}, H. and {Zeilinger}, W.~W. and {Hau}, G.~K.~T.},
        title = "{Dwarf elliptical galaxies with kinematically decoupled cores}",
      journal = {\aap},
         year = 2004,
        month = oct,
       volume = {426},
        pages = {53-63},
          doi = {10.1051/0004-6361:20041205},
archivePrefix = {arXiv},
       eprint = {astro-ph/0407425},
 primaryClass = {astro-ph},
       adsurl = {https://ui.adsabs.harvard.edu/abs/2004A&A...426...53D}
}

@article{Chilingarian2007,
       author = {{Chilingarian}, I.~V. and {Prugniel}, P. and {Sil'Chenko}, O.~K. and {Afanasiev}, V.~L.},
        title = "{Kinematics and stellar populations of the dwarf elliptical galaxy IC 3653}",
      journal = {\mnras},
         year = 2007,
        month = apr,
       volume = {376},
       number = {3},
        pages = {1033-1046},
          doi = {10.1111/j.1365-2966.2007.11549.x},
archivePrefix = {arXiv},
       eprint = {astro-ph/0701842},
 primaryClass = {astro-ph},
       adsurl = {https://ui.adsabs.harvard.edu/abs/2007MNRAS.376.1033C}
}

@article{Thomas2006,
       author = {{Thomas}, Daniel and {Davies}, Roger L.},
        title = "{Rejuvenation of spiral bulges}",
      journal = {\mnras},
         year = 2006,
        month = feb,
       volume = {366},
       number = {2},
        pages = {510-520},
          doi = {10.1111/j.1365-2966.2005.09871.x},
archivePrefix = {arXiv},
       eprint = {astro-ph/0512415},
 primaryClass = {astro-ph},
       adsurl = {https://ui.adsabs.harvard.edu/abs/2006MNRAS.366..510T}
}

@article{Conselice2001,
       author = {{Conselice}, Christopher J. and {Gallagher}, III, John S. and {Wyse}, Rosemary F.~G.},
        title = "{Galaxy Populations and Evolution in Clusters. I. Dynamics and the Origin of Low-Mass Galaxies in the Virgo Cluster}",
      journal = {\apj},
         year = 2001,
        month = oct,
       volume = {559},
       number = {2},
        pages = {791-811},
          doi = {10.1086/322373},
archivePrefix = {arXiv},
       eprint = {astro-ph/0105492},
 primaryClass = {astro-ph},
       adsurl = {https://ui.adsabs.harvard.edu/abs/2001ApJ...559..791C}
}

@article{Conselice2003,
       author = {{Conselice}, Christopher J. and {O'Neil}, Karen and {Gallagher}, John S. and {Wyse}, Rosemary F.~G.},
        title = "{Galaxy Populations and Evolution in Clusters. IV. Deep H I Observations of Dwarf Elliptical Galaxies in the Virgo Cluster}",
      journal = {\apj},
         year = 2003,
        month = jul,
       volume = {591},
       number = {1},
        pages = {167-184},
          doi = {10.1086/375216},
archivePrefix = {arXiv},
       eprint = {astro-ph/0303185},
 primaryClass = {astro-ph},
       adsurl = {https://ui.adsabs.harvard.edu/abs/2003ApJ...591..167C}
}

@article{Sandage1985,
       author = {{Sandage}, A. and {Binggeli}, B. and {Tammann}, G.~A.},
        title = "{Studies of the Virgo cluster. V. Luminosity functions of Virgo cluster galaxies.}",
      journal = {\aj},
         year = 1985,
        month = sep,
       volume = {90},
        pages = {1759-1771},
          doi = {10.1086/113875},
       adsurl = {https://ui.adsabs.harvard.edu/abs/1985AJ.....90.1759S}
}

@article{BinggeliJerjen1998,
       author = {{Binggeli}, Bruno and {Jerjen}, Helmut},
        title = "{Is the shape of the luminosity profile of dwarf elliptical galaxies an useful distance indicator?}",
      journal = {\aap},
         year = 1998,
        month = may,
       volume = {333},
        pages = {17-26},
          doi = {10.48550/arXiv.astro-ph/9704027},
archivePrefix = {arXiv},
       eprint = {astro-ph/9704027},
 primaryClass = {astro-ph},
       adsurl = {https://ui.adsabs.harvard.edu/abs/1998A&A...333...17B}
}

@article{Fillingham2018,
       author = {{Fillingham}, Sean P. and {Cooper}, Michael C. and {Boylan-Kolchin}, Michael and {Bullock}, James S. and {Garrison-Kimmel}, Shea and {Wheeler}, Coral},
        title = "{Environmental quenching of low-mass field galaxies}",
      journal = {\mnras},
         year = 2018,
        month = jul,
       volume = {477},
       number = {4},
        pages = {4491-4498},
          doi = {10.1093/mnras/sty958},
archivePrefix = {arXiv},
       eprint = {1802.03017},
 primaryClass = {astro-ph.GA},
       adsurl = {https://ui.adsabs.harvard.edu/abs/2018MNRAS.477.4491F}
}

@article{RomeroGomez2023,
       author = {{Romero-G{\'o}mez}, J. and {Peletier}, Reynier F. and {Aguerri}, J.~A.~L. and {Mieske}, Steffen and {Scott}, Nicholas and {Bland-Hawthorn}, Joss and {Bryant}, Julia J. and {Croom}, Scott M. and {Eftekhari}, F. Sara and {Falc{\'o}n-Barroso}, Jes{\'u}s and {Hilker}, Michael and {van de Ven}, Glenn and {Venhola}, Aku},
        title = "{The SAMI-Fornax Dwarfs Survey - III. Evolution of [{\ensuremath{\alpha}}/Fe] in dwarfs, from Galaxy Clusters to the Local Group}",
      journal = {\mnras},
         year = 2023,
        month = jun,
       volume = {522},
       number = {1},
        pages = {130-150},
          doi = {10.1093/mnras/stad953},
archivePrefix = {arXiv},
       eprint = {2303.16020},
 primaryClass = {astro-ph.GA},
       adsurl = {https://ui.adsabs.harvard.edu/abs/2023MNRAS.522..130R}
}

@article{Wetzel2015,
       author = {{Wetzel}, Andrew R. and {Deason}, Alis J. and {Garrison-Kimmel}, Shea},
        title = "{Satellite Dwarf Galaxies in a Hierarchical Universe: Infall Histories, Group Preprocessing, and Reionization}",
      journal = {\apj},
         year = 2015,
        month = jul,
       volume = {807},
       number = {1},
          eid = {49},
        pages = {49},
          doi = {10.1088/0004-637X/807/1/49},
archivePrefix = {arXiv},
       eprint = {1501.01972},
 primaryClass = {astro-ph.GA},
       adsurl = {https://ui.adsabs.harvard.edu/abs/2015ApJ...807...49W}
}

@article{Laigle2015,
       author = {{Laigle}, C. and {Pichon}, C. and {Codis}, S. and {Dubois}, Y. and {Le Borgne}, D. and {Pogosyan}, D. and {Devriendt}, J. and {Peirani}, S. and {Prunet}, S. and {Rouberol}, S. and {Slyz}, A. and {Sousbie}, T.},
        title = "{Swirling around filaments: are large-scale structure vortices spinning up dark haloes?}",
      journal = {\mnras},
         year = 2015,
        month = jan,
       volume = {446},
       number = {3},
        pages = {2744-2759},
          doi = {10.1093/mnras/stu2289},
archivePrefix = {arXiv},
       eprint = {1310.3801},
 primaryClass = {astro-ph.CO},
       adsurl = {https://ui.adsabs.harvard.edu/abs/2015MNRAS.446.2744L}
}

@article{Kubo2007,
       author = {{Kubo}, Jeffrey M. and {Stebbins}, Albert and {Annis}, James and {Dell'Antonio}, Ian P. and {Lin}, Huan and {Khiabanian}, Hossein and {Frieman}, Joshua A.},
        title = "{The Mass of the Coma Cluster from Weak Lensing in the Sloan Digital Sky Survey}",
      journal = {\apj},
         year = 2007,
        month = dec,
       volume = {671},
       number = {2},
        pages = {1466-1470},
          doi = {10.1086/523101},
archivePrefix = {arXiv},
       eprint = {0709.0506},
 primaryClass = {astro-ph},
       adsurl = {https://ui.adsabs.harvard.edu/abs/2007ApJ...671.1466K}
}

@article{Bullock2017,
       author = {{Bullock}, James S. and {Boylan-Kolchin}, Michael},
        title = "{Small-Scale Challenges to the {\ensuremath{\Lambda}}CDM Paradigm}",
      journal = {\araa},
         year = 2017,
        month = aug,
       volume = {55},
       number = {1},
        pages = {343-387},
          doi = {10.1146/annurev-astro-091916-055313},
archivePrefix = {arXiv},
       eprint = {1707.04256},
 primaryClass = {astro-ph.CO},
       adsurl = {https://ui.adsabs.harvard.edu/abs/2017ARA&A..55..343B}
}

@article{Lovell2014,
       author = {{Lovell}, Mark R. and {Frenk}, Carlos S. and {Eke}, Vincent R. and {Jenkins}, Adrian and {Gao}, Liang and {Theuns}, Tom},
        title = "{The properties of warm dark matter haloes}",
      journal = {\mnras},
         year = 2014,
        month = mar,
       volume = {439},
       number = {1},
        pages = {300-317},
          doi = {10.1093/mnras/stt2431},
archivePrefix = {arXiv},
       eprint = {1308.1399},
 primaryClass = {astro-ph.CO},
       adsurl = {https://ui.adsabs.harvard.edu/abs/2014MNRAS.439..300L}
}

@article{Vogelsberger2016,
       author = {{Vogelsberger}, Mark and {Zavala}, Jes{\'u}s and {Cyr-Racine}, Francis-Yan and {Pfrommer}, Christoph and {Bringmann}, Torsten and {Sigurdson}, Kris},
        title = "{ETHOS - an effective theory of structure formation: dark matter physics as a possible explanation of the small-scale CDM problems}",
      journal = {\mnras},
         year = 2016,
        month = aug,
       volume = {460},
       number = {2},
        pages = {1399-1416},
          doi = {10.1093/mnras/stw1076},
archivePrefix = {arXiv},
       eprint = {1512.05349},
 primaryClass = {astro-ph.CO},
       adsurl = {https://ui.adsabs.harvard.edu/abs/2016MNRAS.460.1399V}
}

@article{Weisz2014,
       author = {{Weisz}, Daniel R. and {Dolphin}, Andrew E. and {Skillman}, Evan D. and {Holtzman}, Jon and {Gilbert}, Karoline M. and {Dalcanton}, Julianne J. and {Williams}, Benjamin F.},
        title = "{The Star Formation Histories of Local Group Dwarf Galaxies. II. Searching For Signatures of Reionization}",
      journal = {\apj},
         year = 2014,
        month = jul,
       volume = {789},
       number = {2},
          eid = {148},
        pages = {148},
          doi = {10.1088/0004-637X/789/2/148},
archivePrefix = {arXiv},
       eprint = {1405.3281},
 primaryClass = {astro-ph.GA},
       adsurl = {https://ui.adsabs.harvard.edu/abs/2014ApJ...789..148W}
}

@article{Jeon2017,
       author = {{Jeon}, Myoungwon and {Besla}, Gurtina and {Bromm}, Volker},
        title = "{Connecting the First Galaxies with Ultrafaint Dwarfs in the Local Group: Chemical Signatures of Population III Stars}",
      journal = {\apj},
         year = 2017,
        month = oct,
       volume = {848},
       number = {2},
          eid = {85},
        pages = {85},
          doi = {10.3847/1538-4357/aa8c80},
archivePrefix = {arXiv},
       eprint = {1702.07355},
 primaryClass = {astro-ph.GA},
       adsurl = {https://ui.adsabs.harvard.edu/abs/2017ApJ...848...85J}
}
\end{multicols}

\end{document}